\documentclass[nofootinbib,prd,twocolumn,showpacs,showkeys,preprintnumbers]{revtex4-1}
\usepackage{hyperref,amssymb,amsmath,mathrsfs,bm,graphicx}
\begin{document}
\title {Axially symmetric static sources: A general framework and some analytical solutions }
\author{L. Herrera}
\email{laherrera@cantv.net.ve}
\altaffiliation{Also at U.C.V., Caracas}
%\affiliation{Departamento   de F\'{\i}sica Te\'orica e Historia de la  Ciencia,
%Universidad del Pa\'{\i}s Vasco, Bilbao, Spain}
\author{A. Di Prisco}
\email{adiprisc@fisica.ciens.ucv.ve}
%\affiliation{Departamento de F\'\i sica Te\'orica e Historia de la Ciencia,
%Universidad del Pa\'{\i}s Vasco, Bilbao, Spain}
\altaffiliation{Also at U.C.V., Caracas}
\author{J. Ib\'a\~nez  }
\email{j.ibanez@ehu.es}
\affiliation{Departamento de F\'\i sica Te\'orica e Historia de la Ciencia,
Universidad del Pa\'{\i}s Vasco, Bilbao, Spain}
\author{J. Ospino}
\email{jhozcrae@usal.es}
\affiliation{Departamento de Matem\'atica Aplicada,  Universidad de Salamanca, Salamanca, Spain.}
\date{\today}
\begin{abstract}
We provide all basic equations and concepts required to carry out a general study  on axially symmetric static sources. The Einstein equations and the conservation equations are written down for a general anisotropic static fluid endowed with axial symmetry. The structure scalars  are calculated and the inhomogeneity factors are identified. Finally some exact analytical solutions were found. One of these solutions  describes an incompressible spheroid with isotropic pressure and becomes the well known interior Schwarzschild solution in the spherically symmetric limit, however it cannot be matched smoothly to any Weyl exterior metric. Another family of solutions was found that corresponds to an anisotropic fluid distribution and can in principle be matched to a Weyl exterior.
 \end{abstract}
\date{\today}
\maketitle

\section{Introduction}
Observational evidence seems to suggest  that deviations from spherical symmetry in compact
self-gravitating objects (white dwarfs, neutron stars), are
likely to be incidental rather than basic features of these systems.  This explains why spherical symmetry is so commonly assumed in the study of self-gravitating compact objects.

However, the situation is not so simple. Indeed (putting aside the evident fact that astrophysical objects are generally endowed with angular momentum, and therefore excluding all stationary sources), it is well known that the only regular static and asymptotically flat vacuum spacetime posesing a regular horizon is the Schwarzchild solution \cite{israel}, and all the others Weyl exterior solutions \cite{weyl1}-\cite{weyln} exhibit singularities in the  curvature invariants (as the boundary of the source approaches the horizon). This in turn implies that, for  very compact objects, a bifurcation appears between any finite perturbation of Schwarzschild spacetime and any Weyl solution, even when the latter is characterized by parameters arbitrarily close to those corresponding to spherical symmetry (see \cite{i2}-\cite{in} and references therein for a discussion on this point).

From the above comments  it should be clear that a rigorous description of static axially symmetric sources, including finding exact analytical solutions, is a praiseworthy endeavour.

Accordingly, in this work we provide all ingredients and equations required for such a   study. Of course, this issue has already been considered by several authors in the past. Without  pretending to be exhaustive in the revision of  the literature on this problem, let us  mention the pioneering  paper by Hernandez  Jr. \cite{1}, where a general method for obtaining solutions describing axially symmetric sources is presented. Such a method, or some of its modifications  were used in \cite{3} - \cite{7}, to find sources of different Weyl spacetimes.

This problem has also been considered in  \cite{2}-\cite{r3}. However in all these last references the  line element has been assumed to satisfy the so called Weyl gauge, which of course severely restricts the family of  possible sources (see the next section).

In this work we present a general description of axially symmetric sources by deploying all relevant  equations without resorting to the Weyl gauge, and considering the most general matter content consistent with the symmetries of the problem.

With this purpose in mind it would be useful to introduce  the so called  structure scalars. These form a set of scalar functions obtained from the orthogonal splitting of the Riemann tensor. They were originally  defined in the discussion about the structure and evolution of spherically symmetric fluid distributions.  Such scalars (five in the spherically symmetric case) were shown to be endowed with distinct physical meaning \cite{1cil, 2cil, 3cil, 4cil} .

In particular  they control  inhomogeneities in the energy density \cite{1cil}, and the evolution of the expansion scalar and the shear tensor   \cite{1cil, 2cil, 3cil,4cil}. Also  in the static case all possible anisotropic solutions are determined by two structure scalars \cite{1cil}.

Furthermore, the role of electric charge and cosmological constant in  structure scalars has also been recently investigated \cite{5cil}.

More recently, such scalars and their applications were discussed also in the context of cylindrical \cite{scc1}, \cite{scc2} and planar symmetry \cite{csp}.

A set of differential equations for some of these scalars allows us to identify the inhomogeneity factors.

Finally we exhibit two families of solutions. One of them corresponds to an incompressible spheroid with isotropic pressure. It cannot be matched smoothly to any Weyl exterior spacetime. The second one corresponds to an anisotropic fluid and in principle is matchable to a Weyl exterior.

Our paper is organized as follows: In the next section we shall describe the line element corresponding to the most general non--vacuum,  axially symmetric static  spacetime. Next we provide a full description of the source that is represented by a general anisotropic matter. Einstein equations and conservation equations are explicitly written down for such a system.
We also calculate the electric part of the Weyl tensor (its magnetic part vanishes) as well as the electric and magnetic part of the Riemann tensor. With this information we are able to obtain all the non--vanishing structure scalars corresponding to our problem. Two differential equations for such scalars allow us to identify the inhomogeneity factors. The two families of solutions found are described in sections V and VI. A summary of the obtained results as well as a list of some  unsolved issues are presented in section VII . Finally an appendix with the expressions for the components of the electric Weyl tensor, is included.

\section{The metric and the source}
We shall consider bounded, static and axially symmetric sources. For such a system the most general line element may be written in cylindrical coordinates as:

\begin{equation}
ds^2=-A^2 dt^2 + B^2 \left[(dx^1)^2 +(dx^2)^2\right]+D^2d\phi^2,
\label{1}
\end{equation}
where $A, B, D$ are positive functions of $x^1$ and $x^2$. We number the coordinates $x^0=t, x^1=\rho, x^2= z, x^3=\phi$.

We shall  work in ``Weyl spherical coordinates''  ($r,\theta)$ defined by:
\begin{equation}
\rho=r \sin \theta, \qquad z=r\cos\theta.
\label{1coor}
\end{equation}

In these  coordinates the line element reads:
\begin{equation}
ds^2=-A^2 dt^2 + B^2 \left(dr^2 +r^2d\theta^2\right)+D^2d\phi^2,
\label{1b}
\end{equation}
It is important not to confound these coordinates  with Erez--Rosen coordinates $(\hat r, \hat \theta)$ given by
\begin{equation}
\rho^2=(\hat r^2-2m\hat r) \sin^2 \hat \theta, \qquad z=(\hat r-m)\cos\hat \theta,
\label{1coorER}
\end{equation}
where $m$ is a constant to be identified  with the monopole of the source.

It should be stressed  that our line element is defined by three independent functions, unlike  the vacuum case where it is always possible to reduce the line element so that only two independent metric functions appear. In the interior  this is not possible in general, though obviously one may assume that as an additional restriction (the so called Weyl gauge), which amounts to assume that $R^3_3+R^0_0=0$.

In our notation the Weyl gauge is expressed by
\begin{equation}
D=\frac{r \sin \theta}{A}.
\label{wg}
\end{equation}

Let us now provide a full description of the source. In order to give physical significance to the components of the energy momentum tensor, we shall  apply the Bondi approach \cite{Bo}.

Thus, following Bondi, let us introduce purely locally
Minkowski coordinates ($\tau, x, y, z$) (or equivalently,  consider a tetrad
field attached to such l.M.f.) by:
\begin{equation}
d\tau=Adt;\qquad dx=Bdr;\qquad dy=Br d\theta;\qquad dz=Dd\phi.
\label{2}
\end{equation}

Denoting by a hat  the components of the energy--momentum tensor in such locally defined coordinate system, we have that the matter content is given by

\begin{equation}
\widehat{T}_{\alpha\beta}= \left(\begin{array}{cccc}\mu    &  0  &   0     &   0    \\0 &  P_{xx}    &   P_{xy}     &   0    \\0       &   P_{yx} & P_{yy}  &   0    \\0       &   0       &   0     &   P_{zz}\end{array} \right) \label{3},
\end{equation}

where $\mu, P_{xy}, P_{xx}, P_{yy}, P_{zz}$ denote the energy density and different stresses, respectively, as measured by our locally defined Minkowskian observer.

Also observe that  $P_{xy}= P_{yx} $ and, in general  $ P_{xx}  \neq  P_{yy}  \neq P_{zz}$.

Introducing
\begin{equation}
\hat V_\alpha=(-1,0,0,0);\quad \hat K_\alpha=(0,1,0,0);\quad  \hat L_\alpha=(0,0,1,0),
\label{4}
\end{equation}
we have
\begin{eqnarray}
\widehat{T}_{\alpha\beta}&=& (\mu+P_{zz})\hat V_\alpha \hat V_\beta+P_{zz} \eta _{\alpha \beta} +(P_{xx}-P_{zz})\hat K_\alpha \hat K_\beta\nonumber \\ &+& (P_{yy}-P_{zz})\hat L_\alpha \hat L_\beta +2P_{xy}\hat K_{(\alpha} \hat L_{\beta)}
\label{5},
\end{eqnarray}
where $\eta_{\alpha \beta}$ denotes the Minkowski metric.

Then transforming back to our coordinates, we obtain the components of the energy momentum tensor in terms of the physical variables as defined in the l.M.f.
\begin{eqnarray}
{T}_{\alpha\beta}&=& (\mu+P_{zz}) V_\alpha V_\beta+P_{zz} g _{\alpha \beta} +(P_{xx}-P_{zz}) K_\alpha  K_\beta\nonumber \\ &+& (P_{yy}-P_{zz}) L_\alpha L_\beta +2P_{xy} K_{(\alpha}  L_{\beta)},
\label{6}
\end{eqnarray}
where
\begin{equation}
 V_\alpha=(-A,0,0,0);\quad  K_\alpha=(0,B,0,0);\quad  L_\alpha=(0,0,Br,0),
\label{7}
\end{equation}
Alternatively we may write the energy momentum tensor in the ``canonical'' form:
\begin{eqnarray}
{T}_{\alpha\beta}&=& (\mu+P) V_\alpha V_\beta+P g _{\alpha \beta} +\Pi_{\alpha \beta},
\label{6bis}
\end{eqnarray}
with
\begin{eqnarray}
\Pi_{\alpha \beta}&=&(P_{xx}-P_{zz})(K_\alpha K_\beta-\frac{h_{\alpha \beta}}{3})\nonumber \\&+&(P_{yy}-P_{zz})(L_\alpha L_\beta-\frac{h_{\alpha \beta}}{3})+2P_{xy}K_{(\alpha}L_{\beta)}
\label{6bb},
\end{eqnarray}
and
\begin{equation}
P=\frac{P_{xx}+P_{yy}+P_{zz}}{3}, \quad h_{\mu \nu}=g_{\mu \nu}+V_\nu V_\mu.
\label{7P}
\end{equation}
With the above information we can write the Einstein equations, which read:
\begin{widetext}
\begin{eqnarray}
8\pi\mu=-\frac{1}{B^2}\left\{\frac{B^{\prime \prime}}{B}+\frac{D^{\prime \prime}}{D}+\frac{1}{r}(\frac{B^\prime}{B} +\frac{D^\prime}{D})-(\frac{B^\prime}{B})^2+\frac{1}{r^2}\left[\frac{B_{\theta \theta}}{B}+\frac{D_{\theta \theta}}{D}-(\frac{B_\theta}{B})^2\right] \right\},
\label{24}
\end{eqnarray}
\begin{eqnarray}
8\pi P_{xx}=\frac{1}{B^2}\left[\frac{A^\prime B^\prime}{AB}+ \frac{A^\prime D^\prime}{AD}+\frac{B^\prime D^\prime}{BD}+\frac{1}{r}(\frac{A^\prime}{A}+\frac{D^\prime}{D})+\frac{1}{r^2}(\frac{A_{\theta \theta}}{A}+\frac{D_{\theta \theta}}{D}-\frac{A_\theta B_\theta}{AB}+\frac{A_\theta D_\theta}{AD}-\frac{B_\theta D_\theta}{BD})\right],
\label{25}
\end{eqnarray}
\begin{eqnarray}
8\pi P_{yy}=\frac{1}{B^2}\left[\frac{A^{\prime \prime}}{A}+ \frac{D^{\prime \prime}}{D}-\frac{A^\prime B^\prime}{AB} +\frac{A^\prime D^\prime}{AD}-\frac{B^\prime D^\prime}{BD}+\frac{1}{r^2}(\frac{A_\theta B_\theta}{AB}+\frac{A_\theta D_\theta}{AD}+\frac{B_\theta D_\theta}{BD})\right],
\label{27}
\end{eqnarray}
\begin{eqnarray}
8\pi P_{zz}=\frac{1}{B^2}\left\{\frac{A^{\prime \prime}}{A}+ \frac{B^{\prime \prime}}{B}-(\frac{B^\prime}{B})^2+\frac{1}{r}(\frac{A^\prime}{A} +\frac{B^\prime}{B}) +\frac{1}{r^2}\left[\frac{A_{\theta \theta}}{A}+\frac{B_{\theta \theta}}{B}-(\frac{B_\theta}{B})^2\right]\right\},
\label{28}
\end{eqnarray}
\begin{eqnarray}
8\pi P_{xy}=\frac{1}{B^2}\left\{  \frac{1}{r}\left[-\frac{A^{\prime}_\theta}{A} -\frac{D^{\prime}_\theta}{D} +\frac{B_\theta}{B}\left(\frac{A^\prime}{A}+\frac{D^\prime}{D}\right)+\frac{B^\prime}{B}\frac{A_\theta}{A}+\frac{B^\prime}{B}\frac{D_\theta}{D}\right]+\frac{1}{r^2} (\frac{A_\theta}{A}+\frac{D_\theta}{D})\right\},\label{26}
\end{eqnarray}
\end{widetext}
where prime  and subscript $\theta$ denote derivatives with respect to $r$  and $\theta$ respectively.

Also, the nonvanishing components of the conservation equations $T^{\alpha  \beta}_{;\beta}=0$ yield:
\begin{equation}
\dot \mu=0,
\label{21}
\end{equation}
where the overdot denotes derivative with respect to $t$,
and
\begin{widetext}
\begin{eqnarray}
P^\prime_{xx}+\frac{A^{\prime}}{A}(\mu+P_{xx})+\frac{B^{\prime}}{B}(P_{xx}-P_{yy})+\frac{D^{\prime}}{D}(P_{xx}-P_{zz})\nonumber \\+\frac{1}{r}\left[\left(\frac{A_\theta}{A}+2\frac{B_\theta}{B}+\frac{D_\theta}{D}\right)P_{xy}+P_{xy,\theta}+P_{xx}-P_{yy}\right]=0,
\label{22}
\end{eqnarray}
\end{widetext}
\begin{widetext}
\begin{eqnarray}
P_{yy,\theta}+\frac{A_{\theta}}{A}(\mu+P_{yy})+\frac{B_{\theta}}{B}(P_{yy}-P_{xx})+\frac{D_{\theta}}{D}(P_{yy}-P_{zz})+r\left[\left(\frac{A^{\prime}}{A}+2\frac{B^{\prime}}{B}+\frac{D^{\prime}}{D}\right)P_{xy}+P^{\prime}_{xy}\right]+2P_{xy}=0.
\label{23}
\end{eqnarray}
\end{widetext}

Equation (\ref{21}) is a trivial consequence of the staticity, whereas (\ref{22}) and (\ref{23}) are the hydrostatic equilibrium equations.
\section{the structure scalars}
We shall calculate here the structure scalars for the static axially symmetric case. For that purpose, let us first obtain the electric part of the Weyl tensor (the magnetic part vanishes identically).

The components of the electric Weyl tensor can be obtained directly from its definition, 
\begin{equation}
E_{\mu\nu}=C_{\mu\alpha\nu\beta}\,V^\alpha\, V^\beta,\label{8}
\end{equation}
where $C_{\mu\alpha\nu\beta}$ denotes the Weyl tensor. These are exhibited in the Appendix.

Equivalently, the electric part of the Weyl tensor may also be written as:
\begin{widetext}
\begin{eqnarray}
E_{\alpha \beta}=\mathcal{E}_1\left(K_\alpha L_\beta+L_\alpha K_\beta\right)
+\mathcal{E}_2\left(K_\alpha K_\beta-\frac{1}{3}h_{\alpha \beta}\right)+\mathcal{E}_3\left(L_\alpha L_\beta-\frac{1}{3}h_{\alpha \beta}\right), \label{13}
\end{eqnarray}
\end{widetext}
where explicit expressions for the three scalars $\mathcal{E}_1$, $\mathcal{E}_2$, $\mathcal{E}_3$ are given in the Appendix.

Next, let us calculate the electric part of the Riemann tensor (the magnetic part vanishes identically), which is defined by
\begin{equation}
Y^\rho_\beta=V^\alpha V^\mu R^\rho_{\alpha \beta \mu}.
\label{29}
\end{equation}

After some lengthy calculations we find;

\begin{eqnarray}
Y_{\alpha \beta}&=&Y_{TF_1}\left(K_\alpha L_\beta+K_\beta L_\alpha\right)
+Y_{TF_2}\left(K_\alpha K_\beta-\frac{1}{3}h_{\alpha \beta}\right)\nonumber \\
&+&Y_{TF_3}\left(L_\alpha L_\beta-\frac{1}{3}h_{\alpha \beta}\right)+\frac{1}{3} Y_T h_{\alpha \beta},
\label{30}
\end{eqnarray}
where
\begin{equation}
Y_T=4\pi(\mu+P_{xx}+P_{yy}+P_{zz}),
\label{31}
\end{equation}

\begin{equation}
Y_{TF_1}=\mathcal{E}_1-4\pi P_{xy},
\label{32}
\end{equation}
\begin{equation}
Y_{TF_2}=\mathcal{E}_2-4\pi (P_{xx}-P_{zz}),
\label{33}
\end{equation}

\begin{equation}
Y_{TF_3}=\mathcal{E}_3-4\pi (P_{yy}-P_{zz}).
\label{34}
\end{equation}
Finally, we shall find the tensor associated with the double dual of Riemann tensor, defined as:
\begin{equation}
X_{\alpha \beta}=^*R^{*}_{\alpha \gamma \beta \delta}V^\gamma
V^\delta=\frac{1}{2}\eta_{\alpha\gamma}^{\quad \epsilon
\rho}R^{*}_{\epsilon \rho\beta\delta}V^\gamma V^\delta,
\label{35}
\end{equation}
with $R^*_{\alpha \beta \gamma \delta}=\frac{1}{2}\eta
_{\epsilon \rho \gamma \delta} R_{\alpha \beta}^{\quad \epsilon
\rho}$.
Thus, we find
\begin{eqnarray}
X_{\alpha \beta}&=&X_{TF_1}\left(K_\alpha L_\beta+K_\beta L_\alpha\right)+
X_{TF_2}\left(K_\alpha K_\beta-\frac{1}{3}h_{\alpha \beta}\right)\nonumber \\
&+&X_{TF_3}\left(L_\alpha L_\beta-\frac{1}{3}h_{\alpha \beta}\right)+\frac{1}{3}X_T h_{\alpha \beta},
\label{36}
\end{eqnarray}
where
\begin{equation}
X_T=8\pi \mu,
\label{37}
\end{equation}

\begin{equation}
X_{TF_1}=-(\mathcal{E}_1+4\pi P_{xy}),
\label{38}
\end{equation}
\begin{equation}
X_{TF_2}=-\left[\mathcal{E}_2+4\pi (P_{xx}-P_{zz})\right],
\label{39}
\end{equation}

\begin{equation}
X_{TF_3}=-\left[\mathcal{E}_3+4\pi (P_{yy}-P_{zz})\right].
\label{40}
\end{equation}

The  scalars $Y_T$, $Y_{TF1}$, $Y_{TF2}$,$ Y_{TF3}$, $X_T$, $X_{TF1}$, $X_{TF2}$, $X_{TF3}$, are the structure scalars for our problem.

\section{Differential equations for the structure scalars and the inhomogeneity factors}
Two differential equations for the Weyl tensor  may be obtained using Bianchi identities \cite{ellis1}, \cite{ellis2}, they have been found before for the spherically symmetric and the cylindrically symmetric cases (see \cite{scc1}, \cite{inh1} and references therein). Here we calculate them for our case. We obtain:
\begin{widetext}
\begin{eqnarray}
\frac{{\cal E}_{1\theta}}{r}&+&\frac{1}{3}(2{\cal E}_2-{\cal E}_3)^\prime+\frac{{\cal E}_1}{r}\left(\frac{2B_\theta}{B}+
\frac{D_\theta}{D}\right)+{\cal E}_2\left(\frac{B^\prime}{B}+\frac{D^\prime}{D}+\frac{1}{r}\right) -{\cal E}_3 \left(\frac {B^\prime}{B}+\frac{1}{r}\right)=
\frac{4\pi}{3}\left(2\mu+P_{xx}+P_{yy}+P_{zz}\right)^\prime\\ \nonumber
&+&4\pi(\mu+P_{xx})\frac{A^\prime}{A}+4\pi P_{xy}\frac{A_\theta}{Ar},
\label{55}
\end{eqnarray}
\end{widetext}
\begin{widetext}
\begin{eqnarray}
{\cal E}^{\prime}_1&+&\frac{1}{3r}(2{\cal E}_3-{\cal E}_2)_\theta+{\cal E}_1\left(\frac{2B^\prime}{B}+
\frac{D^\prime}{D}+\frac{2}{r}\right)-\frac{{\cal E}_2 B_\theta}{Br}+\frac{{\cal E}_3}{r} \left(\frac{B_\theta}{B}+\frac{D_\theta}{D}\right)=
\frac{4\pi}{3r}\left(2\mu+P_{xx}+P_{yy}+P_{zz}\right)_\theta\\ \nonumber &+&4\pi(\mu+P_{yy})\frac{A_\theta}{Ar}
+4\pi P_{xy}\frac{A^\prime}{A},
\label{56}
\end{eqnarray}
\end{widetext}
which, using (\ref{31})-(\ref{34}) and (\ref{37})-{\ref{40}), may be written in terms of structure scalars:
\begin{widetext}
\begin{eqnarray}
\frac{8 \pi \mu^\prime}{3}=-\frac{1}{r}\left[X_{TF1\theta}+X_{TF1}(\ln {B^2 D)}_{\theta}
\right]-\left[\frac{2}{3}X_{TF2}^\prime+X_{TF2}(\ln{BDr})^{\prime}\right]
+\left[\frac{1}{3}X_{TF3}^\prime+X_{TF3}(\ln{Br})^{\prime}\right],
\label{57}
\end{eqnarray}
\end{widetext}
\begin{widetext}
\begin{eqnarray}
\frac{8 \pi \mu_{\theta}}{3r}=\frac{1}{r}\left[\frac{1}{3}X_{TF2\theta}+X_{TF2}(\ln B)_{\theta}\right]-\frac{1}{r}
\left[\frac{2}{3}X_{TF3\theta}+X_{TF3}(\ln{ BD})_{\theta}\right]-
\left[X_{TF1}^\prime+X_{TF1}(\ln{B^2 D r^2})^{\prime}\right].
\label{58}
\end{eqnarray}
\end{widetext}

Let us now turn to the inhomogeneity factors.
The inhomogeneity factors (say $\Psi_i$) are the specific combinations of physical and geometric  variables,
such  that  their vanishing is a necessary and sufficient condition for the homogeneity of energy
density (i.e. for the vanishing of all spatial derivatives of the energy density).

In the spherically symmetric case it has been shown that in the absence of dissipation the necessary and sufficient condition for the vanishing of the (invariantly defined) spatial derivative of  the energy density  is the vanishing of the scalar  associated with the trace free part of  $X_{\alpha \beta}$ (see \cite{1cil}, \cite{inh1}).

We shall now identify the inhomogeneity factors in our case.

First, observe that from (\ref{57}) and (\ref{58}) it follows at once that $X_{TF1}=X_{TF2}=X_{TF3}=0\Rightarrow  \mu^\prime=\mu_\theta=0$. In order to identify the above scalars as the inhomogeneity factors we  need to  prove that the inverse is also true (i.e. $\mu^\prime=\mu_\theta=0\Rightarrow X_{TF1}=X_{TF2}=X_{TF3}=0$).

For that purpose we shall first establish the behaviour of different variables in the neighborhood of $r\approx 0$.
We shall demand that both $A$ and $B$ are regular functions, and
\begin{equation}
 D(r,\theta)\approx r\sin\theta,
\label{nr}
\end{equation}
at $r=0$.
Then from (\ref{22}) it follows that in the neighborhood of $r=0$
\begin{equation}
P_{xy}\approx r,\quad P_{xx}-P_{yy}\approx r,\quad P_{xx}-P_{zz} \approx r,
\label{59}
\end{equation}
and from (\ref{26})
\begin{equation}
A_\theta(0,\theta)\approx r^3,\qquad B_\theta(0,\theta)\approx r^3,\qquad A(0,\theta)^\prime_\theta \approx r^2.
\label{60}
\end{equation}

Using the above in (\ref{18}) we have that
\begin{equation}
{\cal E}_1(0,\theta)\approx r,
\label{61}
\end{equation}
implying because of (\ref{38})
\begin{equation}
X_{TF1}(0,\theta)\approx r.
\label{62}
\end{equation}

Finally we shall assume that the three structure scalars $X_{TF1}, X_{TF2}, X_{TF3}$ are analytical functions (class $C^\infty$) in the neighborhood of  $r=0$.

Then, if we  assume $\mu^\prime=\mu_\theta=0$.  Evaluating  (\ref{58}) in the neighborhood of $r\approx 0$,
since $X^\prime_{TF1}(r=0)$ is regular,  it follows
\begin{equation}
X^\prime_{TF1}(r=0)=0,
\label{63}
\end{equation}
and
\begin{equation}
X_{TF2\theta}-2 X_{TF3\theta}-X_{TF3}\cot\theta=0,
\label{64}
\end{equation}
where (\ref{nr}), (\ref{60}) and (\ref{63}) have been used (in what follows it is understood that all expressions are evaluated at $r\approx0$).

Next, from (\ref{57}) we obtain
\begin{equation}
2X_{TF2}=X_{TF3},
\label{65}
\end{equation}
where (\ref{63}) and the regularity of first derivatives of structure scalars have been used.

Then, feeding back (\ref{65}) into (\ref{64}) we obtain
\begin{equation}
X_{TF2}=\frac{\alpha}{\sin^{2/3}\theta},
\label{66}
\end{equation}
where $\alpha$ is a constant, implying $X_{TF2}=X_{TF3}=0$.

Thus in the neigborhood of $r\approx 0$ we have $X_{TF1}=X^\prime _{TF1}=X_{TF2}=X_{TF3}=0$.

Next, taking $r$-derivative of (\ref{58}), and evaluating at $r\approx 0$ we get $X^{\prime \prime}_{TF1}=0$. Continuing this process it follows that $X^{(n)}_{TF1}\equiv \frac{\partial^nX_{TF1}}{\partial r^n}=0$ for any $n \geq 0$.

Also, from the $r$-derivative of (\ref{57}) evaluated in the neighborhood of $r\approx 0$ it follows
\begin{equation}
2 X^\prime_{XTF2}=X^\prime_{TF3}.
\label{67}
\end{equation}
Feeding this equation back into the $r$-derivative of (\ref{58}) produces
\begin{equation}
X^\prime_{TF2}=\frac{\beta}{\sin^{2/3}\theta},
\label{68}
\end{equation}
where $\beta$ is a constant, implying $X^\prime_{TF2}=X^\prime_{TF3}=0$.
It is not difficult to see that this procedure can be continued to obtain, in the neighborhood of $r\approx 0$, $X^{(n)} _{TF1}=X^{(n)}_{TF2}=X^{(n)}_{TF3}=0$, for any $n\geq 0$. Therefore we can continue analytically  their value at the center, from which we infer:
\begin{equation}
X_{TF1}=X_{TF2}=X_{TF3}=0\Leftrightarrow  \mu^\prime=\mu_\theta=0.
\label{70}\end{equation}

This last result allows us to identify the three structure scalars $X_{TF1}, X_{TF2}, X_{TF3}$ as the inhomogeneity factors.

We shall next find some explicit analytical solutions.

\section{the incompressible, isotropic  spheroid}
We shall now find an analytical solution corresponding to a bounded spheroid with isotropic pressures and homogeneous energy density.  From the results of the previous section, (\ref{32})-(\ref{34}) and (\ref{38})-(\ref{40}), it is evident that such a solution is also conformally flat.

Thus let us assume $P_{xx}=P_{yy}=P_{zz}=P$, $P_{xy}=0$ and $\mu=\mu_0=constant$

For simplicity we shall assume the boundary surface $\Sigma$ to be defined by the equation:
\begin{equation}
r=r_1=constant.
\label{50}
\end{equation}
Then to  satisfy Darmois conditions (continuity of the first and second fundamental forms) we demand  that all metric functions as well as $r$ derivatives, to be continuous across $\Sigma$ (see \cite{1}). Obviously $\theta$ derivatives of $A, B, D$ and $A^\prime, B^\prime, D^\prime$ are continuous too across $\Sigma$.

From  the above and  (\ref{25}) and (\ref{26}) it follows that
\begin{equation}
P\stackrel{\Sigma}{=}0.
\label{51}
\end{equation}
Under the conditions above (\ref{22}) and (\ref{23}) can be integrated to obtain:
\begin{equation}
P+\mu_0=\frac{\zeta}{A},
\label{52}
\end{equation}
and
\begin{equation}
P+\mu_0=\frac{\xi(r)}{A},
\label{53}
\end{equation}
where $\xi$ is  an arbitrary function of its argument.
Using boundary conditions (\ref{51}) in (\ref{52}) (\ref{53}) it follows that:
\begin{equation}
 A(r_1,\theta)=const.=\frac{\alpha}{\mu_0}, \qquad \zeta=constant.
\label{54}
\end{equation}

Since, as mentioned before, our solution is conformally flat, then using $\mathcal{E}_1=\mathcal{E}_2=\mathcal{E}_3=0$ and
$P_{xy}=0,$ in  (\ref{26}) and (\ref{15}) we obtain
\begin{equation}
\frac{A^\prime _\theta}{A}-\frac{A^\prime }{A}\frac{B
_\theta}{B}-\frac{A _\theta}{A}\left (\frac{B^\prime
}{B}+\frac{1}{r}\right )=0,\label{26c}
\end{equation}
\begin{equation}
\frac{D^\prime _\theta}{D}-\frac{D^\prime }{D}\frac{B
_\theta}{B}-\frac{D _\theta}{D}\left (\frac{B^\prime
}{B}+\frac{1}{r}\right )=0.\label{15c}
\end{equation}

Introducing the auxiliary function $\bar A(r,\theta)$ defined by
\begin{equation}
A(r,\theta)=\bar A(r,\theta)B(r,\theta),
\label{red1}
\end{equation}
and assuming
\begin{equation}
D(r,\theta)=B(r,\theta)r \sin \theta,\label{cfs}
\end{equation}
the equations (\ref{26c}) and (\ref{15c}) can be integrated to
obtain

\begin{eqnarray}
\bar A(r,\theta)&=&\tilde A(r) +r \chi(\theta),\nonumber \\
B(r,\theta)&=&\frac{1}{R(r)+r \omega (\theta)},\label{acsol}
\end{eqnarray}
where $\tilde A$, $\chi$, $R$ and $\omega$ are arbitrary functions of  their argument.

Next, from (\ref{16}) and (\ref{17}), taking into account
(\ref{cfs}) and (\ref{acsol}) we get

\begin{equation}
\chi= a\cos\theta,\quad  \quad \tilde A(r)=\alpha
r^2+\beta.\label{omegaB}
\end{equation}

Where $a$, $\alpha$ and $\beta$ are constants of integration.

 From the above it follows that conformally flat solutions are
 described by the line element
\begin{widetext}
 \begin{equation}
 ds^2=\frac{1}{\left[R(r)+r \omega (\theta) \right]^2}\left [-(\alpha r^2+\beta+a r\cos\theta)^2dt^2+dr^2+r^2 d\theta ^2+r^2\sin ^2 (\theta)d\phi ^2\right ]. \label{cfm}
\end{equation}
\end{widetext}

Next, from the condition $P_{yy}-P_{zz}=0$ and the equations
(\ref{27})-(\ref{28}) we get

\begin{equation}
\omega (\theta)=b\cos\theta,\label{p2}
\end{equation}

and from the condition $P_{xx}-P_{yy}=0$ and the equations
(\ref{25})-(\ref{26}) we obtain
\begin{equation}
R(r)=\gamma r^2+\delta, \label{Rcf}
\end{equation}

where $b$, $\gamma$ and $\delta$ are constants of integration.

Finally, the metric of incompressible conformally flat isotropic
fluids can be written as follows.

\begin{widetext}
 \begin{equation}
 ds^2=\frac{1}{(\gamma r^2+\delta +b r\cos \theta)^2}\left [-(\alpha r^2+\beta+a r\cos\theta)^2dt^2+dr^2+r^2d\theta ^2+r^2\sin ^2 \theta d\phi ^2\right ]. \label{cfifm}
\end{equation}
\end{widetext}

Next,  the physical variables can be easily calculated.  Thus, using (\ref{cfifm}) into (\ref{24}) the energy density reads:

 \begin{equation}
8\pi \mu = 12 \gamma \delta-3b^2.
\label{den1}
\end{equation}

To obtain  the pressure we shall use (\ref{52}) and (\ref{54}), which produce
\begin{equation}
8\pi P =(3b^2-12\gamma\delta )\left [1-\frac{\alpha
r_1^2+\beta}{\gamma r_1^2+\delta} \frac{\gamma
r^2+\delta+br\cos\theta}{\alpha r^2+\beta +a
r\cos\theta}\right ],\label{epf}
\end{equation}
where  

\begin{equation}
\zeta=\mu_0\frac{\alpha r_1^2+\beta}{\gamma
r_1^2+\delta},\quad  a=\frac{\alpha r_1^2+\beta}{\gamma
r_1^2+\delta}b,
\label{jcn}
\end{equation}
in order to satisfy  the junction condition  (\ref{51}).

It may be instructive to recover the spherically symmetric case (the interior Schwarzschild solution). In this case we have $a=b=0$.

To see how this comes about, let us perform the transformation
\begin{equation}
\bar r=\frac{r}{\gamma r^2+\delta},\qquad \bar \theta=\theta,\qquad \bar t=t,\qquad \bar \phi=\phi,
\label{esf1}
\end{equation}
where $overbar$ denotes the usual Schwarzschild  coordinates.
Then, it is a simple matter to check that  (\ref{cfifm}) and (\ref{epf})  are identical to  the well known expressions characterizing the interior Schwarzschild solution:

\begin{equation}
g_{\bar t \bar t}=\frac{1}{4}\left[3(1-\frac{2M}{\bar r_1})^{(1/2)}-(1-\frac{2m(\bar r)}{\bar r})^{(1/2)}\right],
\label{esf3}
\end{equation}

\begin{equation}
g_{\bar r \bar r}=(1-\frac{2m(\bar r)}{\bar r})^{-1},
\label{esf4}
\end{equation}

\begin{equation}
P=\mu\left[\frac{(1-\frac{2m(\bar r)}{\bar r})^{1/2}-(1-\frac{2M}{\bar r_1})^{1/2}}{3(1-\frac{2M}{\bar r_1})^{1/2}-(1-\frac{2m(\bar r)}{\bar r})^{1/2}}\right],
\label{esf5}
\end{equation}
\\
where $m(\bar r)$, $M$ and $\bar r_1$ denote the mass function, the total mass and the radius of the sphere respectively,
and  the following relationships are satisfied
\begin{equation}
m=\frac{4 \pi}{3}\mu \bar r^3=\frac{2\gamma \delta r^3}{(\gamma r^2+\delta)^3},
\label{esf6}
\end{equation}
\begin{equation}
m(\bar r_1)=M,
\label{esf6bis}
\end{equation}
and
\begin{equation}
r_1^2=\frac{\beta + \delta }{\gamma-\alpha}.
\label{esf7}
\end{equation}

At this point it is pertinent to ask the question: to what specific exterior spacetime can we match smoothly our solution?
This is a relevant question, since there are
as many different (physically distinguishable) Weyl solutions as there are different
harmonic functions.

The answer to the above question is the following: our solution cannot be matched to any Weyl exterior, even though it  has a surface of vanishing pressure. This is so because  the first  fundamental form is not continuous across the boundary surface. 

Indeed  from the continuity of $g_{tt}$ and $g_{\phi \phi}$ components at $r=r_1$, we have:
\begin{equation}
A_W(r_1,\theta)=\frac{\alpha r_1^2+\beta+a r_1\cos\theta}{\gamma r_1^2+\delta +b r_1\cos \theta},
\label{imm}
\end{equation}
and
\begin{equation}
A_W(r_1,\theta)=\gamma r_1^2+\delta +b r_1\cos \theta,
\label{immII}
\end{equation}
where $A_W(r_1,\theta)$ denotes the $\sqrt{g_{tt}}$ of any Weyl exterior solution  (evaluated on the boundary surface).
It is a simple matter to check that the two  equations above  cannot be satisfied unless $a=b=0$ which corresponds to the spherically symmetric case. This result is in agreement with  theorems indicating that static, perfect fluid (isotropic in pressure) sources are spherical (see \cite{prueba} and references therein)

Observe that the  above result is a consequence of (\ref{cfs}). Therefore in order to  find matchable solutions we should relax this condition. In this later case, of course, neither isotropy of pressure nor homogeneity of the energy density is preserved.

The remaining possibility is trying to match on a boundary surface given by the equation $r=r_1(\theta)$. However this does not seem to solve the problem since in our case $A(r_1,\theta)=constant$.

\section{Anisotropic inhomogeneous spheroids}
In order to find solutions that could be matched to a Weyl exterior, we shall relax the condition of isotropy of pressure and energy density homogeneity.

The introduction of the pressure anisotropy is well motivated from purely physical considerations. Indeed, it is well known that  compact objects (white dwarfs and neutron stars)  are endowed with strong magnetic fields (see \cite{mag1}--\cite{mag7} and references therein). On the other hand it has been shown that the effect of such a magnetic field  on a degenerate Fermi gas manifests itself through the appearance of  a strong anisotropy  due to the magnetic stresses (see \cite{anis1}--\cite{anis4} and references therein). This in turn may  severely affect some important characteristics of compact objects (see \cite{ef1}--\cite{ef3} and references therein). Besides the magnetic field, local anisotropy of pressure may be produced  by a variety of physical phenomena (see \cite{rev} and references therein).

Thus,  let us assume $P_{xy}=\mathcal{E}_1=\mathcal{E}_3=P_{yy}-P_{zz}=0$, although  $\mathcal{E}_2\neq 0$ and $P_{xx}\neq P_{yy}$. Then  from the
equation(\ref{26}) and (\ref{18}) we obtain the equations
(\ref{26c}) and (\ref{15c}).

Next, introducing the auxiliary
functions $\tilde A(r,\theta) $ and $R(r,\theta)$ defined by
\begin{eqnarray}
A(r,\theta)=\tilde A(r,\theta) B(r,\theta)r,\nonumber \\
D(r,\theta)=R(r,\theta) B(r,\theta)r,\label{AR}
\end{eqnarray}
the equation (\ref{26c}) and (\ref{15c}) can be rewritten as:
\begin{equation}
\frac{\tilde A^\prime _\theta}{\tilde A}=\frac{R^\prime
_\theta}{R}=Br(\frac{1}{Br})^\prime_\theta.\label{o1}
\end{equation}
From  $P_{yy}-P_{zz}=\mathcal
E_3=0$ and the equations (\ref{17}) (\ref{20}) we get
\begin{widetext}
 \begin{equation}
 \frac{R^{\prime\prime}}{R}+\frac{R^\prime}{R}(\frac{1}{r}-\frac{\tilde A^\prime}{\tilde A})+\frac{1}{r^2}(\frac{\tilde A_{\theta\theta}}{\tilde A}-\frac{\tilde A_\theta}{\tilde A}\frac{R_\theta}{R})=0,
 \end{equation}
 \begin{equation}
 \frac{R^\prime}{R}(\frac{\tilde A^\prime
 }{\tilde A}+\frac{B^\prime}{B}+\frac{1}{r})+\frac{1}{r^2}(\frac{\tilde
 A_\theta}{\tilde
 A}\frac{R_\theta}{R}+\frac{B_\theta}{B}\frac{R_\theta}{R}+2\frac{B^2_\theta}{B^2}-\frac{\tilde
 A_{\theta\theta}}{\tilde
 A}-\frac{B_{\theta\theta}}{B})=0.\label{o3}
 \end{equation}
 \end{widetext}
In order to find a simple solution  that satifies  the boundary
condition, we choose:
 \begin{equation}
R(r,\theta)=\sin \theta  \left[b(r_1)+\gamma
\cos \theta \right ],
\label{R}
\end{equation} then from the equations
(\ref{o1})-(\ref{o3}) we find
\begin{equation}
\tilde A(r,\theta)=\alpha\left[\frac{\gamma}{2} \sin^2\theta-b(r_1)\right]
\cos\theta+ a(r),\label{oA}
\end{equation}

\begin{equation}
B(r,\theta)=\frac{1}{\beta r\left [\frac{\gamma}{2} \sin^2 \theta-b(r_1)\cos\theta\right]+ b(r)},\label{oA}
\end{equation}
\\

where $\alpha, \gamma, \beta$ and $a(r), b(r)$ are arbitrary constants and functions of integration respectively.
$a(r)$, $\alpha$ and  $\beta$  have the dimension of an inverse of length, whereas the others are dimensionless.

For the above metric, Einstein equations yield the following expressions for physical variables.
\begin{widetext}
\begin{eqnarray}
8\pi \mu&=&(\beta r \Gamma+b)^2\left\{\frac{2b^{\prime \prime}}{\beta r \Gamma+b}-\frac{3(\beta \Gamma+b^{\prime})^2}{(\beta r \Gamma+b)^2}-\frac{3(\beta \Lambda \sin \theta )^2}{(\beta r \Gamma+b)^2}+\frac{1}{r}\left[\frac{4(\beta \Gamma +b^{\prime})}{\beta r \Gamma+b}+\frac{4\beta \Sigma}{\beta r \Gamma+b}\right]-\right.\nonumber \\& -&\left.\frac{1}{r^2}\left[1-\frac{b_1+4\gamma \cos \theta}{\Lambda}\right] \right\},
\label{dns}
\end{eqnarray}
\end{widetext}

\begin{widetext}
\begin{eqnarray}
8\pi P_{xx}&=&(\beta r \Gamma+b)^2 \left\{\frac{-2a^{\prime}(\beta \Gamma+b^{\prime})}{(\beta r \Gamma+b)(\alpha \Gamma+a)}+\frac{3(\beta \Gamma+b^{\prime})^2}{(\beta r \Gamma+b)^2}+\frac{3(\beta \Lambda\sin \theta )^2}{(\beta r \Gamma+b)^2}\right. \nonumber \\&+& \left.\frac{1}{r}\left[\frac{2a^{\prime}}{\alpha \Gamma +a}-\frac{6(\beta \Gamma +b^{\prime})}{\beta r \Gamma+b}-\frac{4\beta\Sigma}{\beta r \Gamma +b}-\frac{2\alpha \beta \Lambda^2\sin^2 \theta }{(\alpha \Gamma+a)(\beta r \Gamma+b)}\right]\right. \nonumber \\&-& \left.\frac{1}{r^2}\left[3+\frac{2\alpha\Sigma}{\alpha \Gamma+a}-\frac{b_1+4\gamma \cos \theta}{\Lambda}\right] \right\},
\label{pxx}
\end{eqnarray}
\end{widetext}

\begin{widetext}
\begin{eqnarray}
8\pi P_{yy}=8\pi P_{zz}&=&(\beta r \Gamma+b)^2 \left\{\frac{a^{\prime \prime}}{\alpha \Gamma+a}-\frac{2b^{\prime \prime}}{\beta r \Gamma+b}-\frac{2 a^{\prime}(\beta \Gamma+b^{\prime})}{(\beta r \Gamma+b)(\alpha \Gamma+a)}+\frac{3(\beta \Gamma+b^{\prime})^2}{(\beta r \Gamma+b)^2}+\frac{3(\beta \Lambda\sin \theta )^2}{(\beta r \Gamma+b)^2}\right. \nonumber \\&+& \left.\frac{1}{r}\left[\frac{3a^{\prime}}{\alpha \Gamma +a}-\frac{4(\beta \Gamma +b^{\prime})}{\beta r \Gamma+b}-\frac{2\beta \Sigma}{\beta r \Gamma +b}-\frac{2\alpha \beta \Lambda^2\sin^2 \theta }{(\alpha \Gamma+a)(\beta r \Gamma+b)}\right]+\frac{1}{r^2}\left(1+\frac{\alpha \Sigma}{\alpha \Gamma+a} \right) \right\},
\label{pyy}
\end{eqnarray}
\end{widetext}
where
\begin{eqnarray}
\Lambda&=&b_1+\gamma \cos \theta,  \qquad \Gamma=\frac{\gamma \sin^2 \theta}{2}-b_1 \cos \theta, \qquad b(r_1)=b_1\nonumber\\ \Sigma&=&\Lambda \cos \theta -\gamma \sin^2 \theta.
\label{int}
\end{eqnarray}
The  equations above describe a wide class of solutions that can be matched to any specific Weyl metric by an appropriate choice of functions and constants of integration. Furthermore physically reasonable  models can be obtained at least from slight deviations from spherical symmetry.
\section{conclusions}
We have established the general framework to carry out a systematic analysis of general static axially symmetric sources. By ``general'' we mean that the Weyl gauge was not assumed and the matter description is the most general compatible with axial symmetry and staticity.

Thus,  we started from the most general line element and  considered  a general anisotropic fluid as source of the exterior Weyl spacetime. Relevant equations were then written down and structure scalars were calculated.

We have seen that the three structure scalars associated with the tracefree part of the tensor $X_{\alpha \beta}$ define the inhomogeneity factors.

We have found an exact analytical solution representing a spheroid of isotropic pressure with homogeneous energy density. In the spherically symmetric limit our solution becomes the well known Schwarzschild interior solution. Such an interior cannot be matched (except in the spherically symmetric case) to any Weyl exterior.

Matchable solutions can be found by relaxing the conditions of isotropy of pressure and density inhomogeneity. The anisotropy of pressure was also justified on physical grounds. An example was given in the last section. In order to study the physical relevance of nonsphericity in the structure of the source it is necessary to match the above metioned solution to a specific Weyl exterior (so that the arbitrary parameters of the source could be related to the parameters of the exterior metric), but such a task   is beyond the scope of this paper.

In general, solutions as the one presented here and many others found by either analytical or numerical procedures could provide answers to important questions related to stellar structure, namely:
\begin{itemize}
\item What is the limit of compactness of a static axially symmetric source?
\item How is the above limit related to (influenced by) some exterior parameters such as the quadrupole moment of the source?
\item How are,  intrinsically nonspherical physical variables ( e.g. $P_{xy}$), related to multipole moments (higher than monopole)?
\item Are specific Weyl exteriors related to specific sources?
\end{itemize}
\begin{acknowledgments}

L.H. thanks  Departamento   de F\'{\i}sica Te\'orica e Historia de
la  Ciencia, Universidad del Pa\'{\i}s Vasco  for financial
support and hospitality. ADP and J.O.  acknowledge hospitality of
the Departamento   de F\'{\i}sica Te\'orica e Historia de la
Ciencia, Universidad del Pa\'{\i}s Vasco. This work was partially
supported by the Spanish Ministry of Science and Innovation (grant
FIS2010-15492) and UFI 11/55 program of the Universidad del Pa\'\i
s Vasco. J.O. acknowledges financial support from the Spanish
Ministry of Science and Innovation (grant FIS2009-07238).
\end{acknowledgments}
\appendix*
\section{Expression for the components of the electric Weyl tensor} 
The nonvanishing components as calculated from (\ref{8}) are:
\begin{widetext}
\begin{eqnarray}
E_{11} &=& \frac{1}{6}\left[\frac{2A^{\prime \prime}}{A}-\frac{B^{\prime \prime}}{B}-\frac{D^{\prime \prime}}{D}-\frac{3A^{\prime} B^{\prime}}{AB}-\frac{A^{\prime} D^{\prime}}{AD}+(\frac{B^{\prime}}{B})^2+\frac{3B^{\prime} D^{\prime}}{BD}+\frac{1}{r}(2\frac{D^\prime}{D}-\frac{B^\prime}{B}-\frac{A^\prime}{A})\right]\nonumber \\
&+&\frac{1}{6r^2}\left[-\frac{A_{\theta \theta}}{A} -\frac{B_{\theta \theta}}{B} +\frac{2 D_{\theta \theta}}{D} +\frac{3A_\theta B_\theta}{AB} -\frac{A_\theta D_\theta}{AD}+(\frac{B_\theta}{B})^2-\frac{3B_\theta D_\theta}{BD}\right],\label{9}
\end{eqnarray}
\end{widetext}
\begin{widetext}

\begin{eqnarray}
E_{22} &=& -\frac{r^2}{6}\left[\frac{A^{\prime \prime}}{A}+\frac{B^{\prime \prime}}{B}-\frac{2D^{\prime \prime}}{D}-\frac{3A^{\prime} B^{\prime}}{AB}+\frac{A^{\prime} D^{\prime}}{AD}-(\frac{B^{\prime}}{B})^2+\frac{3B^{\prime} D^{\prime}}{BD}+\frac{1}{r}(\frac{D^\prime}{D}+\frac{B^\prime}{B}-\frac{2A^\prime}{A})\right]\nonumber \\
&-&\frac{1}{6}\left[-\frac{2A_{\theta \theta}}{A} +\frac{B_{\theta \theta}}{B} +\frac{ D_{\theta \theta}}{D} +\frac{3A_\theta B_\theta}{AB} +\frac{A_\theta D_\theta}{AD}-(\frac{B_\theta}{B})^2-\frac{3B_\theta D_\theta}{BD}\right],\label{10}
\end{eqnarray}
\end{widetext}
\begin{widetext}
\begin{eqnarray}
E_{33} &=& -\frac{D^2}{6B^2}\left[\frac{A^{\prime \prime}}{A}-\frac{2B^{\prime \prime}}{B}+\frac{D^{\prime \prime}}{D}-\frac{2A^{\prime} D^{\prime}}{AD}+2(\frac{B^{\prime}}{B})^2+\frac{1}{r}(\frac{D^\prime}{D}-\frac{2B^\prime}{B}+\frac{A^\prime}{A})\right]\nonumber \\
&-&\frac{D^2}{6B^2r^2}\left[\frac{A_{\theta \theta}}{A} -\frac{2B_{\theta \theta}}{B} +\frac{ D_{\theta \theta}}{D}  -\frac{2A_\theta D_\theta}{AD}+2(\frac{B_\theta}{B})^2\right],\label{11}
\end{eqnarray}
\end{widetext}
\begin{widetext}
\begin{eqnarray}
E_{12}  = \frac{1}{2} \left[\frac{A^{\prime}_\theta}{A} -\frac{D^{\prime}_\theta}{D}+\frac{B_\theta}{B}\frac{D^{\prime}}{D}-\frac{A^\prime B_{\theta}}{AB}-\frac{B^\prime A_{\theta}}{AB}+\frac{D_\theta}{D}\frac{B^{\prime}}{B}-\frac{1}{r}\left(\frac{A_\theta}{A}-
\frac{D_\theta}{D}\right)\right] \label{12}.
\end{eqnarray}
\end{widetext}
These components are not independent since they satisfy the relationship:
\begin{equation}
E_{11}+\frac{1}{r^2}E_{22}+\frac{B^2}{D^2}E_{33}=0.
\end{equation}

For the three scalars $\mathcal{E}_1, \mathcal{E}_2, \mathcal{E}_3$ we obtain:

\begin{widetext}
\begin{eqnarray}
\mathcal{E}_1= \frac{1}{2B^2} \left[\frac{1}{r}(\frac{A^{\prime}_\theta}{A} -\frac{D^{\prime}_\theta}{D}-
\frac{B_\theta}{B}\frac{A^{\prime}}{A}+\frac{D^{\prime}}{D}\frac{B_{\theta}}{B}-\frac{B^\prime}{B}\frac{A_\theta}{A}+
\frac{D_\theta}{D}\frac{B^\prime}{B})+\frac{1}{r^2}(\frac{D_{\theta}}{D}-\frac{A_\theta}{A})\right],\label{15}
\end{eqnarray}
\end{widetext}
\begin{widetext}
\begin{eqnarray}
\mathcal{E}_2 & = &-\frac{1}{2B^2}\left[-\frac{A^{\prime \prime}}{A}+\frac{B^{\prime \prime}}{B}+
\frac{A^\prime B^\prime}{AB}+\frac{A^\prime D^\prime}{AD}-(\frac{B^\prime}{B})^2-\frac{B^\prime D^\prime}{BD}+\frac{1}{r}(\frac{B^\prime}{B}
-\frac{D^\prime}{D})\right]\nonumber \\ &-&\frac{1}{2B^2r^2}\left[\frac{B_{\theta \theta}}{B} -
\frac{D_{\theta \theta}}{D} -\frac{A_{\theta}B_{\theta}}{AB} +\frac{A_\theta D_\theta}{AD} -
(\frac{B_\theta}{B})^2+\frac{B_\theta D_\theta}{BD}\right],\label{16}
\end{eqnarray}
\end{widetext}
\begin{widetext}
\begin{eqnarray}
\mathcal{E}_3 & = &-\frac{1}{2B^2}\left[\frac{B^{\prime \prime}}{B}-\frac{D^{\prime \prime}}{D}-
\frac{A^\prime B^\prime}{AB}+\frac{A^\prime D^\prime}{AD}-(\frac{B^\prime}{B})^2+\frac{B^\prime D^\prime}{BD}+\frac{1}{r}(\frac{B^\prime}{B}
-\frac{A^\prime}{A})\right]\nonumber \\ &-&\frac{1}{2B^2r^2}\left[\frac{B_{\theta \theta}}{B} -\frac{A_{\theta \theta}}{A}
 +\frac{A_{\theta}B_{\theta}}{AB} +\frac{A_\theta D_\theta}{AD} -
(\frac{B_\theta}{B})^2-\frac{B_\theta D_\theta}{BD}\right].\label{17}
\end{eqnarray}
\end{widetext}
Or, using Einstein equations we may also write:
\begin{widetext}
\begin{eqnarray}
\mathcal{E}_1 =\frac{E_{12}}{B^2r}=4\pi P_{xy}+\frac{1}{B^2 r}\left[\frac{A^{\prime}_\theta}{A}-
\frac{A^{\prime} B_\theta}{AB}-\frac{A_\theta}{A} (\frac{B^{\prime}}{B}+\frac{1}{r})\right],
\label{18}
\end{eqnarray}
\end{widetext}
\begin{widetext}
\begin{eqnarray}
\mathcal{E}_2 &=&-\frac{2E_{33}}{D^2}-\frac{E_{22}}{B^2r^2}=
{4\pi} (\mu+2P_{xx}+P_{yy})
-\frac{A^\prime}{B^2A}\left(\frac{2D^{\prime}}{D}
+\frac{B^{\prime}}{B}+\frac{1}{r}\right)\nonumber \\
&+&\frac{A_\theta}{AB^2r^2} \left(\frac{B_\theta}{B}-\frac{2D_\theta}{D}\right)-\frac{1}{B^2r^2}\frac{A_{\theta \theta}}{A},
\label{19}
\end{eqnarray}
\end{widetext}
\begin{widetext}
\begin{eqnarray}
\mathcal{E}_3 =-\frac{E_{33}}{D^2}+\frac{E_{22}}{B^2r^2}=4\pi (P_{yy}-P_{zz})
-\frac{A^\prime}{B^2A}\left(\frac{D^{\prime}}{D}
-\frac{B^{\prime}}{B}-\frac{1}{r}\right)
-\frac{A_\theta}{AB^2r^2} \left(\frac{D_\theta}{D}+\frac{B_\theta}{B}\right)+
\frac{1}{B^2r^2}\frac{A_{\theta \theta}}{A}.
\label{20}
\end{eqnarray}
\end{widetext}
 

\begin{thebibliography}{100}
\bibitem{israel} W. Israel {\it Phys. Rev.} {\bf 164}, 1776, (1967).
\bibitem{weyl1} H.  Weyl {\it Ann. Physik} {\bf 54}, 117 (1918,).
\bibitem{weyl2} H. Weyl  {\it Ann. Physik} {\bf 364}, 185 (1919).
\bibitem{weyl3}T. Levi-Civita {\it  Atti. Accad. Naz. Lincei Rend. Classe Sci.Fis. Mat.Nat.} {\bf 28}, 101 (1919).
\bibitem{weyln} J. L. Synge {\it Relativity, The general theory}(North-Holland Publ. Co, Amsterdam) (1960).
\bibitem{i2} J.  Winicour, A.I. Janis and E.T. Newman {\it  Phys. Rev.} {\bf 176},1507 (1968).
\bibitem{i3} A.  Janis, E.T Newman and J. Winicour {\it Phys. Rev. Lett.} {\bf 20}, 878 (1968).
\bibitem{i1} L.  Bel {\it Gen. Relativ. Gravitation} {\bf 1}, 337 (1971).
\bibitem{i4} F. I. Cooperstock and G. J. Junevicus {\it Nuovo Cimento} {\bf 16B}, 387 (1973).
\bibitem{i5} L. Herrera {\it Int. J. Mod. Phys. D} {\bf 17 }, 557 (2008).
\bibitem{in} L. Herrera {\it Int. J. Mod. Phys. D} {\bf 17 }, 2507 (2008).
\bibitem{1}W. C. Hernandez, Jr., {\it Phys. Rev.} {\bf 153}, 1359 (1967).
\bibitem{3}B. W. Stewart, D. Papadopoulos, L. Witten, R. Berezdivin and L. Herrera, {\it Gen. Rel. Grav.} {\bf 14}, 97 (1982).
\bibitem{6} L. Herrera, G. Magli and D. Malafarina {\it Gen. Rel. Grav.} {\bf 37}, 1371 (2005).
\bibitem{7}L. Herrera, W. Barreto  and J. L.  Hern\'andez--Pastora {\it Gen. Rel. Grav.} {\bf 37}, 873 (2005).
\bibitem{2}J. J. J. Marek, {\it Phys. Rev.} {\bf 163}, 1373 (1967).
\bibitem{r1} K. Y. Fu {\it Astrophys. J.} {\bf  190}, 411 (1974).
\bibitem{r2} H. Ardavan and M. Hossein Partovi {\it Phys. Rev. D} {\bf 16}, 1664 (1977).
\bibitem{r3} W. B. Bonnor {\it An interior solution for the Curzon body} (2012).
\bibitem{1cil} L. Herrera, J. Ospino, A. Di Prisco, E. Fuenmayor and  O. Troconis
{\it Phys. Rev. D} {\bf 79}, 064025 (2009).
\bibitem{2cil} L. Herrera, A. Di Prisco, J. Ospino and J. Carot {\it Phys. Rev.D} {\bf 82}, 024021 (2010).
\bibitem{3cil} L. Herrera, A. Di Prisco and J. Ospino {\it Gen.Rel. Grav.} {\bf 42}, 1585 (2010).
\bibitem{4cil}  L. Herrera, A. Di Prisco and   J. Ib\'a\~nez
{\it  Phys. Rev. D} {\bf  84},  064036 (2011).
\bibitem{5cil}  L. Herrera, A. Di Prisco and   J. Ib\'a\~nez
{\it  Phys. Rev. D} {\bf  84}, 107501  (2011).
\bibitem{scc1}  L. Herrera, A. Di Prisco and J. Ospino {\it Gen.Rel. Grav.} {\bf 44}, 2645 (2012).
\bibitem{scc2}  M. Sharif and M. Zaeem Ul Haq Bhatti {\it Gen.Rel. Grav.} {\bf 44},  2811 (2012).
\bibitem{csp} M. Sharif  and M. Zaeem Ul Haq Bhatti {\it Mod. Phys. Lett. A} {\bf 27}, 1250141 (2012).
\bibitem{Bo} H. Bondi, {\it Proc. R. Soc. London}, {\bf A281}, 39 (1964)

\bibitem{ellis1} G. F. R. Ellis {\ Relativistic Cosmology} in: Proceedings of the International School of Physics `` Enrico Fermi'', Course 47: General Relativity and Cosmology. Ed. R. K. Sachs (Academic Press, New York and London) (1971).
\bibitem{ellis2} G. F. R. Ellis {\it Gen. Rel. Grav.} {\bf  41}, 581 (2009).
\bibitem{inh1} L.Herrera {\it Int. J. Mod. Phys. D} {\bf 20}, 1689 (2011).
\bibitem{prueba} A. K. M. Masood-ul-Alama {\it Gen. Rel. Grav.} {\bf 39}, 55 (2007).
\bibitem{mag1} F. Pacini {\it Nature} {\bf 216}, 567 (1967).
\bibitem{mag2} T. Gold {\it Nature} {\bf 218}, 731 (1968).
\bibitem{mag3}J. P.Ostriker and  J. E. Gunn {\it Astrophys. J.} {\bf 157}, 1395 (1969).
\bibitem{mag4} J. C. Kemp, J. B. Swedlund, J. D. Landstreet and J. R. P. Angel {\it Astrophys. J} {\bf 161}, L77 (1970).
\bibitem{mag5} G. D. Schmidt and P. S. Smith {\it Astrophys. J.} {\bf 448}, 305 (1995).
\bibitem{mag6} A. Putney {\it Astrophys. J.} {\bf 451}, L67 (1995).
\bibitem{mag7} D. Reimers, S. jordan, D. Koester, N. Bade, Th. Kohler and L. Wisotzki {\it Astron. Astrophys.} {\bf 311}, 572 (1996).
\bibitem{anis1} V. Canuto and H. Y. Chiu {\it Phys. Rev.} {\bf 173}, 1210 (1968).
\bibitem{anis2} M. Chaichian, S. S. Masood, C. Montone, A. P\'erez-Martinez and H. P\'erez Rojas {\it Phys. Rev. Lett.} {\bf 84} 5261 (2000).
\bibitem{anis3} A. P\'erez-Martinez, H. P\'erez Rojas and H. J. Mosquera Cuesta {\it Int. J. Mod. Phys. D} {\bf17}, 2107 (2008).
\bibitem{anis4} E. J. Ferrera, V.  de la Incera, J. P. Keith, I. Portillo and P. L. Springsteen {\it Phys. Rev. C}  {\bf 82}, 065802 (2010).
\bibitem{ef1} R. Gonz\'alez Felipe and A. P\'erez-Martinez {\it J. Phys. G: Nucl. Part. Phys.} {\bf 36}, 075202 (2009).
\bibitem{ef2} M. Sinha, B. Mukhopadhyay  and A. Sedrakian, {\it Arxiv: 10005.4995v2} (2010).
\bibitem{ef3} U. Das and B.  Mukhopadhyay {\it Int. J. Mod. Phys. D}  {\bf 21}, 1242001 (2010).
\bibitem{rev} L. Herrera and N. O. Santos {\it Phys. Rep.} {\bf 286}, 53 (1997).
\end{thebibliography}
\end{document}